\documentstyle[aps,prb,twocolumn,epsf]{revtex}
\begin{document}
\draft
\twocolumn[\hsize\textwidth\columnwidth\hsize\csname
@twocolumnfalse\endcsname 
\title{Rabi oscillations, coherent properties and model qubits \\in two-level donor systems under terahertz radiation}
\author{H. S. Brandi$^{a,b}$, A. Latg\'e$^{c}$, and L. E. Oliveira$^{d}$}
\address{$^{a}$Instituto de F\'{i}sica, Univ. Federal do Rio de Janeiro, Rio de Janeiro - RJ, 21945-970, Brazil}
\address{$^{b}$Inmetro, Campus de Xer\'{e}m, Duque de Caxias - RJ, 25250-020, Brazil}
\address{$^{c}$Instituto de F\'{i}sica, Univ. Federal Fluminense, 
 Niter\'{o}i - RJ, 24210-340, Brazil}
\address{$^{d}$Instituto de F\'{i}sica, Unicamp, CP 6165, Campinas - SP, 13083-970, Brazil}
\date{\today}
\maketitle
\begin{abstract}
Quantum confinement, magnetic-field effects, and laser coupling with
the two low-lying states of electrons bound to donor impurities in
semiconductors may be used to coherently manipulate the two-level
donor system in order to establish the appropriate operational
conditions of basic quantum bits (qubits) for solid-state based
quantum computers. Here we present a detailed theoretical calculation
of the damped Rabi oscillations and time evolution of the 1s and
$2p_+$ donor states in bulk GaAs under an external magnetic field and
in the presence of terahertz laser radiation, and their influence on
the measured photocurrent. We also detail the possible experimental
conditions under which decoherence is weak and qubit operations are
efficiently controlled.
\end{abstract}

]

\narrowtext

Some two decades ago, it was suggested that quantum-mechanical
computers (QC) might provide much more computing power than machines
based on classical physics \cite{Feynman}. More recently, the
discovery of fast quantum algorithms for tasks such as prime
factorization and searching disordered databases {\cite{Shor} as well
as quantum error correction codes \cite{DiV0} have arisen considerable
interest in the search for suitable physical systems on which reliable
quantum hardware could be realized. A precisely defined two-fold
Hilbert space may be regarded as the base of a QC, under the
conditions that decoherence is weak and single-qubit and two-qubit
unitary operations are controlled.

Many implementations have been suggested for quantum computation
involving manipulation of nuclear spin states in bulk solution using
nuclear magnetic resonance techniques (NMR) \cite{Chuang}, trapped
ions \cite{Zoller}, photons in microcavity \cite{Kimble}, and
specifically selected donor \cite{Cole} or exciton
\cite{Cundiff,Kamada,Zrenner} states in semiconductor systems as
qubits to be manipulated by laser radiation. Solid-state proposals
open up the possibility of fabricating large integrated networks that
would be required for realistic applications of quantum computers. Two
interesting suggestions for such semiconductor QCs are based on qubits
using the spin of a phosporus donor nucleus in bulk silicon
\cite{Kane} and the spins of electrons trapped in a GaAs quantum dot
(QD) \cite{DiV,Sarma0}. Of course, a silicon-based computer is
extremely attractive from many aspects but unfortunately difficulties
concerning its practical realization have been pointed out recently
\cite{Koiller}. The main problem is related to the strong oscillations
in the exchange splitting- basic ingredient for the two-qubit
operation in silicon-based QC, originated from degeneracy of the
conduction band minimum in Si. This situation does not occur in GaAs
where coupled-QD structures have been previously theoretically
investigated \cite{DiV,Sarma0,Sarma} considering two coupled GaAs
QDs. This corresponds essentially to ``a 2D - hydrogen-like
molecule'', in the presence of an external magnetic field. An
extensive study by Hu and Das Sarma \cite{Sarma0} and de Sousa {\it et
al} \cite{Sarma} suggested that the two-QD system would provide the
necessary two-qubit entanglement required for QCs. In this case, the
QDs provide the tag for each qubit which are represented by the spin
of the QD-trapped electron, and single-qubit operations are performed
on the tag-spin by a local external magnetic field or equivalently by
an efficient manipulation of the local spin states \cite{Loss}. The
two-qubit operation, which rotates one of the tag-spin qubit in a
precise angle if, and only if, the other control-qubit is oriented in
a well defined unchanged direction, is realized via exchange
interaction between the electrons in each QD. This means that in order
to have a precise control of two-qubit operations it is necessary to
fine-tune the exchange coupling through gates which are physically
realized by external fields. One should note that a QC based on this
architeture would be effective only if the spin decoherence time is
much longer than the time involved in the single- and two-qubit
operations, which could pose a problem, as these operations would be
controlled by switching electric and/or magnetic fields and this
cannot be performed very fast. It is possible to overcome this
limitation if, instead of slow external gate potential and/or magnetic
fields, one uses a laser-probe pulse to control the qubit
operations. This can be achieved via coherent manipulation of
two-level systems through Rabi oscillations in donor \cite{Cole} or
exciton \cite{Cundiff,Kamada,Zrenner} states, by applying
electromagnetic fields. Cole {\it et al} \cite{Cole} and Zrenner {\it
et al} \cite{Zrenner} have demonstrated that the coherent optical
excitations in a QD and bulk GaAs two-level systems, respectively, can
be converted into deterministic photocurrents. Here, the impurities
(dots) provide the tag for each qubit which are represented by the
two-level system. Moreover, it has been shown \cite{Brandi0} that
donor states in QDs, under magnetic and pump-laser \cite{Cohen}
fields, may be properly manipulated in order to produce the required
robust donor states for terahertz coherent manipulation of qubits. In
the present work, we present a detailed theoretical calculation of the
time evolution of the 1s and $2p_+$ donor states in bulk GaAs and
their influence on the photocurrent \cite{Cole}, and thoroughly detail
the conditions under which decoherence is weak.

A set of optical Bloch equations \cite{Cohen} is used to describe the
time evolution of the elements of the density matrix within a
two-level model for the impurity system, i.e.,

\begin{eqnarray}
\nonumber \frac{d\rho_{11}}{dt}&=&-i \Omega_{R} \cos( \omega_{L}t)(\rho_{21}-\rho_{12}) + \gamma_{1}
 \rho_{22}\\
\frac{d\rho_{22}}{dt} &=& +i \Omega_{R} \cos( \omega_{L}t)(\rho_{21}-\rho_{12}) - (\gamma_{1}+\gamma_{3}) \rho_{22}\\
\nonumber \frac{d\rho_{12}}{dt} &=& +i \omega_{21}\rho_{12} + i\Omega_{R} \cos( \omega_{L}t) (\rho_{11}-\rho_{22}) - \gamma_{2}\rho_{12}\\
\nonumber \frac{d\rho_{21}}{dt} &=& -i \omega_{21}\rho_{21} - i\Omega_{R} \cos( \omega_{L}t) (\rho_{11}-\rho_{22}) - \gamma_{2}\rho_{21}
\end{eqnarray}
where $\omega_{L}$ is the THz laser frequency, $\omega_{21}$ is the
energy separation of the 1s and 2p$_+$ impurity levels,
$\Omega_{R}=E_{THz}d_{12}^{x}/\hbar$ is the Rabi frequency, $E_{THz}$
is the amplitude of the terahertz electric field (in the
$\it{x}$-direction), and $d_{12}^{x}$ is the $\it{x}$-component of the
1s-2p$_+$ dipole matrix element \cite{note}. The parameters
$\gamma_{1}$, $\gamma_{2}$, and $\gamma_{3}$ are recombination rates
as defined in Cole {\it et al} \cite{Cole}. The 1s- and 2p$_+$-like
impurity states are calculated by using hydrogenic-like wave functions
with exponentially-decaying variational parameters taken as
$\lambda_{1s}$ and $\lambda_{2p_+}$. The calculated energies, in
effective units, are given by

\begin{equation}
\epsilon_{1s} = E_c + \frac{1}{\lambda^2_{1s}} - \frac{2}{\lambda_{1s}} + \frac{1}{2}\gamma^2 \lambda_{1s}^2 \,\,\\\,,
\end{equation}

\begin{equation}
\epsilon_{2p_+} = E_c + \frac{1}{\lambda^2_{2p_+}} -\frac{1}{\lambda_{2p_+}}+\gamma + \frac{3}{2}\gamma^2 \lambda^2_{2p_+}\,\,,
\end{equation}
where $E_{c} = E_c^o + \gamma$, $E_c$ being the magnetic-field
dependent conduction-band edge, and $\gamma = e\hbar B/2m^*cR^*$ is
the effective magnetic energy, with $R^*\approx 5.9$ $\it {meV}$ the
GaAs effective Rydberg. The parameters $\lambda_{1s}$ and
$\lambda_{2p_+}$ are obtained variationally as a function of the
applied magnetic field (in the $\it{z}$-direction). Fig. 1(a) shows
the energies of both impurity states together with the magnetic-field
dependence of the energy of the conduction-band edge. By using the
value of $\nu = 2.52$ $\it {THz}$ for the radiation frequency as in
the experiment by Cole {\it et al} \cite{Cole}, we calculated the
magnetic field value of 3.4 T for which the 1s-2p$_+$ transition is
tuned with the radiation. This is in good agreement with the
experimental value of 3.51 T obtained by  Cole {\it et al} \cite{Cole}
at very low values of the terahertz electric field. The 1s and 2p$_+$
hydrogenic-like wave functions, for the impurity under a magnetic
field of 3.4 T, are used to theoretically evaluate the Rabi frequency
as a function of the THz electric field, and results are shown as a
full curve in Fig. 1(b). Also shown are the fit parameters by Cole
{\it et al} \cite{Cole} as full triangles. We note that Cole {\it et
al} \cite{Cole} have pointed out that the uncertainty of the absolute
scale of $E_{THz}$ is $\pm$ 50 $\%$ and the uncertainty in relatives
values of $E_{THz}$ is negligible. With this in mind, one observes
that the open triangles, which correspond to a horizontal shift of the
fit parameters by 1.1 x 10$^{4}$ Vm$^{-1}$, are in quite good
agreement with our theoretical prediction for the Rabi frequency.

The equations of motion of the density matrix may be solved via the
rotating-wave scheme \cite{Cohen}, and solutions essentially oscillate
with a frequency close to the Rabi frequency, with damping terms
depending on the various recombination mechanisms. To discuss the
photocurrent measurements in Fig. 4(a) of Cole {\it et al}
\cite{Cole}, we choose $\gamma_1$=0, $\gamma_2 = 1.5$ x $10^{11}$ rad
$s^{-1}$, $\gamma_3 = 1.0 $ x $ 10^{11}$ rad $s^{-1}$, and $\Omega_R =
3.0 $ x $ 10^{11}$ rad $s^{-1}$. Our calculated results for the
photocurrent [taken as proportional to $1 - \rho_{11}(t)$] in
Fig. 2(a) show that this choice of parameters fits quite well their
experimental data for the photocurrent, in the range of measured
detunings. Note that the theoretical photocurrent results in Fig. 2(a)
indicate [cf. Fig. 1(b)] that the experimental data in Fig. 4(a) of
Cole {\it et al} \cite{Cole} should be assigned to a laser amplitude
of $E_{THz}\approx$ 3.1 x 10$^{4}$ Vm$^{-1}$. In Fig. 2(b), we display
the theoretical photocurrent for a laser detuning $\delta$= -0.014 THz
as compared with $\rho_{22}(t)$ (number of donors at the 2$p_+$
state), and $\rho_{11}(t) + \rho_{22}(t)$ (total population in the 1s
and 2$p_+$ states). One notes that the populations of both the 1s and
2$p_+$ states have a fast decrease with pulse duration. Therefore, it
is clear that the measured photocurrent may have a dominant
contribution from the charge ionized from the 2$p_+$ state into the
conduction band, making the existence of coherent Rabi oscillations
more difficult to be observed, {\it i.e.}, the fact that the 2$p_+$
state is immersed in the conduction band results in a strong
decoherence process. It is important to mention here that it is
possible to increase the confinement of the donor electrons by
trapping the impurities in QDs or via the use of a pump-laser tuned
below the semiconductor gap, in order to lower the energy of the
2$p_+$ state and to bring it below the conduction-band continuum \cite
{Brandi0}. This situation corresponds to a choice of the $\gamma_3$
ionization rate equal to zero, as shown for the calculated
photocurrent in Fig. 3(a). Alternatively, one may efficiently control
the decoherence process through a diluted doping procedure in which
the inhomogeneous broadening of the 1s-2$p_+$ transition energy is
conveniently reduced. This occurs because the overlap between the
impurity wave functions diminishes. Again,  the confinement of the
electronic states via a pump-laser field may be used to mimic this
effect. Such behaviour may be modelled by assuming the total dephasing
rate $\gamma_2 = 0$, as displayed in Fig. 3(b). One should note that a
too low impurity concentration may result in a weak photosignal which
is a rather undesirable experimental situation. In the theoretical
photocurrent displayed in Fig. 3(c), it is quite clear the obvious
fundamental role played by the strength of the electromagnetic field,
through the Rabi oscillations, on the coherence process.

In conclusion, we have obtained the solutions of the optical Bloch
equations within the rotating-wave approximation in order to discuss
the proposal of Cole {\it et al} \cite{Cole}. The donor 1s and 2$p_+$
hydrogenic-like states, in the presence of a homogeneous magnetic
field, are obtained through a variational procedure. Theoretical
calculations have shown that the resonance condition - for which the
1s-2$p_+$ transition is tuned with the terahertz radiation - is
achieved for a magnetic field in good agreement with the measured
experimental value. Moreover, the theoretical results obtained for the
THz electric field-dependence of the Rabi frequencies are found in
excellent agreement with the experimentally fitted curve, provided
that the uncertainty of the absolute experimental scale for the
electric field is taken into account. The theoretical calculations
concerning the time-dependent behaviour of the photocurrent, a
fundamental observable for an efficient determination of qubit
operations, show that it is dominated by the ionized charge in the
conduction band, and that this effect relates to the 2$p_+$ state
being immersed in the conduction band. In that respect, we have
detailed the possible experimental conditions for which decoherence
effects are minimized so that qubit operations are efficiently
controlled.


We are grateful to R. R. dos Santos for a critical reading of the manuscript. The authors would like to thank the Brazilian Agencies CNPq, FAPERJ, FUJB, FAPESP, and the MCT - Institute of Millenium for partial financial support. 


\newpage

\begin{figure}[tbp]
\epsfxsize=2.1in 
\centerline{\epsfbox{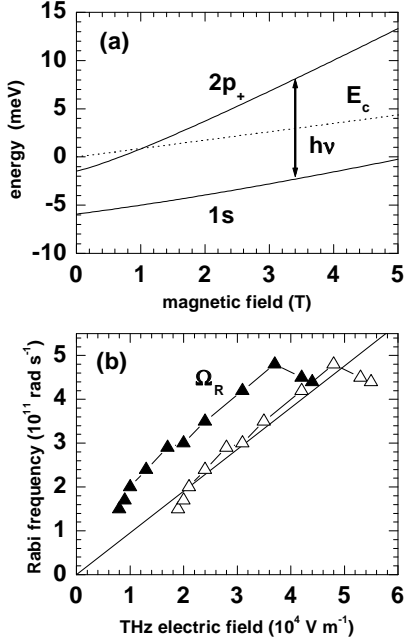}}
\caption{(a) Theoretical magnetic - field dependence of the 1s and 2$p_+$ donor states together with the field-dependence of the GaAs $E_c$ conduction-band edge. At 3.4 T, one obtains a 1s-2$p_+$ transition tunable to an energy corresponding to 2.52 THz; (b) Rabi frequency as a function of the THz electric field (full curve), obtained via a variational calculation (see text). Full triangles are the fit parameters by Cole {\it et al} \cite{Cole}, with open triangles corresponding to a horizontal shift by 1.1 x 10$^{4}$ Vm$^{-1}$.}
\end{figure}

\begin{figure}[tbp]
\epsfxsize=2.1in 
\centerline{\epsfbox{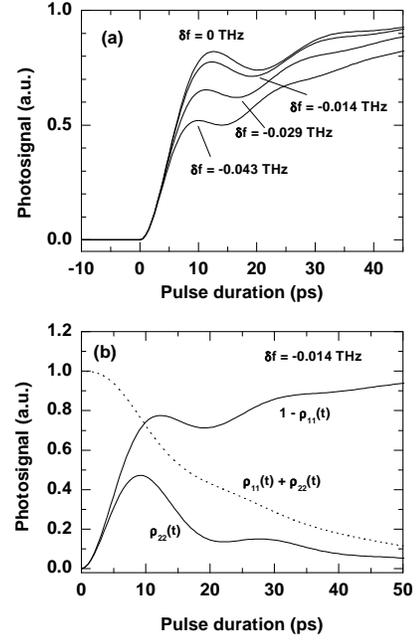}}
\caption{(a) Theoretical $1 - \rho_{11}(t)$ photocurrent (arbitrary units) as a function of pulse duration for various laser detunings $\delta$=0 THz, -0.014 THz, -0.029 THz, and -0.043 THz; (b) Photocurrent for a laser detuning $\delta$= -0.014 THz as compared with $\rho_{22}(t)$ (number of donors at the 2$p_+$ state), and $\rho_{11}(t) + \rho_{22}(t)$. Results are calculated using $\gamma_1$=0, $\gamma_2 = 1.5$ x $10^{11}$ rad $s^{-1}$, $\gamma_3 = 1.0 $ x $ 10^{11}$ rad $s^{-1}$, and $\Omega_R = 3.0 $ x $ 10^{11}$ rad $s^{-1}$. }
\end{figure}

\begin{figure}[tbp]
\epsfxsize=2.7in 
\centerline{\epsfbox{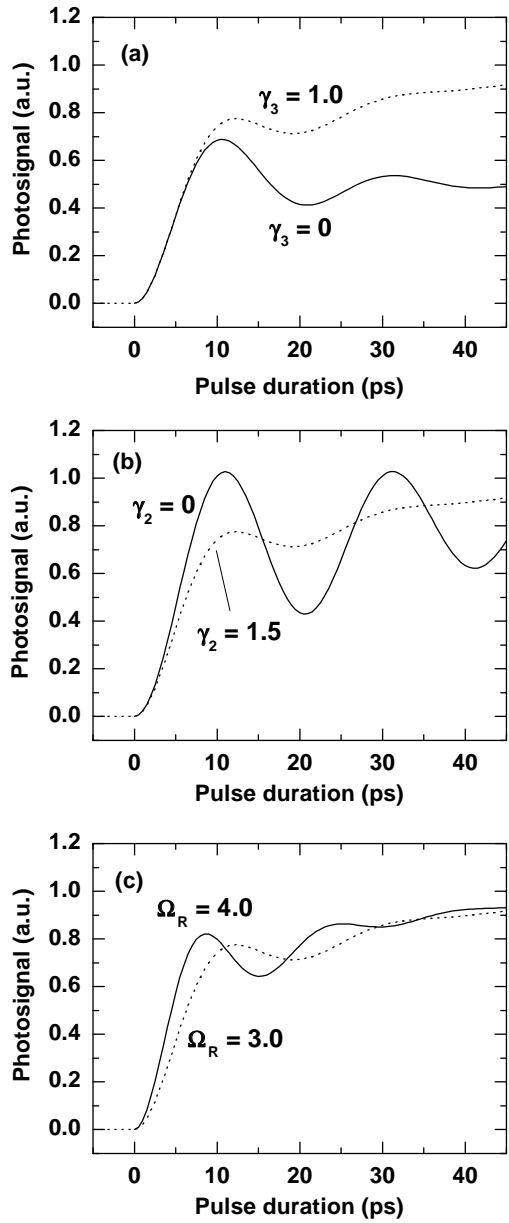}}
\caption{Dependence of the photocurrent (arbitrary units) on the pulse duration for a fixed laser detuning $\delta$ = -0.014 THz and (a) $\gamma_3 = 0$, and $\gamma_3 = 1.0$ (with $\gamma_2 = 1.5$ and $\Omega_R = 3.0 $ ), (b) $\gamma_2 = 0$, and $\gamma_2 = 1.5$ (with $\gamma_3 = 1.0 $ and $\Omega_R = 3.0$); and (c) $\Omega_R = 3.0$, and $\Omega_R = 4.0$ (with $\gamma_2 = 1.5$ and $\gamma_3 = 1.0 $). Rabi frequencies and rate parameters are given in units of $10^{11}$ rad $s^{-1}$.}
\end{figure}

\end{document}